\documentclass[10pt,twocolumn,letterpaper]{article}

\usepackage[pagenumbers]{cvpr} 

\usepackage{graphicx}
\usepackage{amsmath}
\usepackage{amssymb}
\usepackage{booktabs}
\usepackage{multirow}
\usepackage{comment}
\usepackage{enumerate}
\usepackage{enumitem}
\usepackage{hhline}
\usepackage[accsupp]{axessibility}

\usepackage[pagebackref,breaklinks,colorlinks]{hyperref}

\usepackage[capitalize]{cleveref}
\crefname{section}{Sec.}{Secs.}
\Crefname{section}{Section}{Sections}
\Crefname{table}{Table}{Tables}
\crefname{table}{Tab.}{Tabs.}

\def\cvprPaperID{1897}
\def\confName{CVPR}
\def\confYear{2022}

\newcommand{\RED}[1]{#1}

\makeatletter
\def\thanks#1{\protected@xdef\@thanks{\@thanks
        \protect\footnotetext{#1}}}
\makeatother

\begin{document}

\setlength{\abovecaptionskip}{1pt}     
\setlength{\belowcaptionskip}{-8pt}    

\title{Physical Inertial Poser (PIP): Physics-aware Real-time Human Motion Tracking from Sparse Inertial Sensors\thanks{This work was supported by Beijing Natural Science Foundation (JQ19015), the NSFC (No.61727808, 62021002), the National Key R\&D Program of China 2018YFA0704000. This work was supported by THUIBCS, Tsinghua University and BLBCI, Beijing Municipal Education Commission. This work was partially supported by the ERC consolidator grant 4DReply (770784). We thank Notiom~\cite{Noitom} for the extensive support on inertial sensors, and Liuqing Yang, Liangdi Ma, Siyuan Teng, Wenbin Lin for the help on live demos. Feng Xu is the corresponding author. }}  

\author{ 
    Xinyu Yi\textsuperscript{1} \qquad 
    Yuxiao Zhou\textsuperscript{1} \qquad 
    Marc Habermann\textsuperscript{2} \qquad 
    Soshi Shimada\textsuperscript{2} \qquad 
    \\
    Vladislav Golyanik\textsuperscript{2} \qquad 
    Christian Theobalt\textsuperscript{2} \qquad 
    Feng Xu\textsuperscript{1}
    \\
    \small{\textsuperscript{1}School of Software and BNRist, Tsinghua University}
    \quad
    \small{\textsuperscript{2}Max Planck Institute for Informatics, Saarland Informatics Campus} 
}

\maketitle

\begin{abstract}
Motion capture from sparse inertial sensors has shown great potential compared to image-based approaches since occlusions do not lead to a reduced tracking quality and the recording space is not restricted to be within the viewing frustum of the camera.
However, capturing the motion and global position only from a sparse set of inertial sensors is inherently ambiguous and challenging.
In consequence, recent state-of-the-art methods can barely handle very long period motions, and unrealistic artifacts are common due to the unawareness of physical constraints.
To this end, we present the first method which combines a neural kinematics estimator and a physics-aware motion optimizer to track body motions with only 6 inertial sensors.
The kinematics module first regresses the motion status as a reference, and then the physics module refines the motion to satisfy the physical constraints.
Experiments demonstrate a clear improvement over the state of the art in terms of capture accuracy, temporal stability, and physical correctness.

\end{abstract}

\section{Introduction}
\begin{figure}
    \centering
    \includegraphics[width=\linewidth]{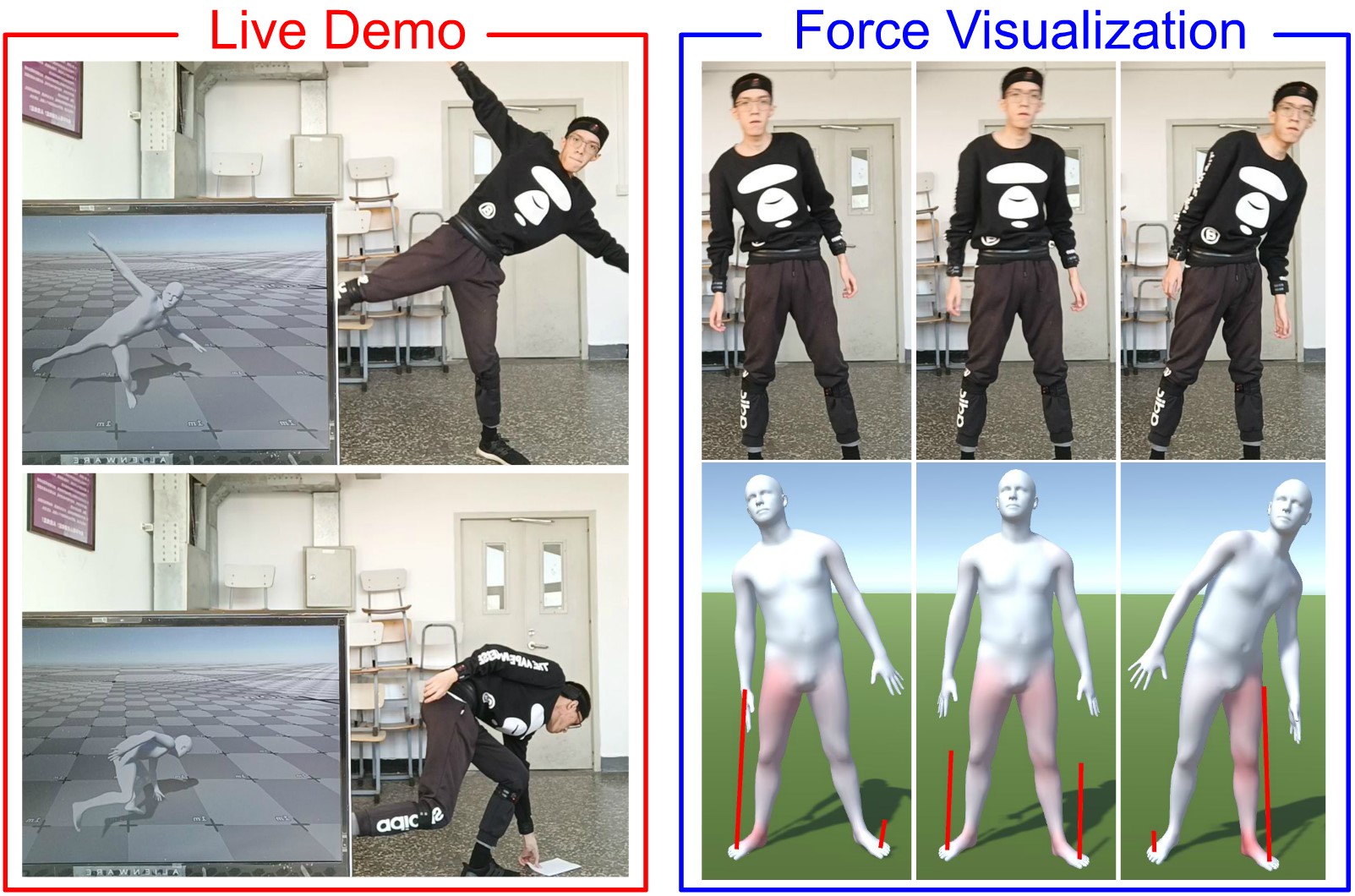}
    \caption{
      \textbf{PIP captures physically correct human motion, joint torques, and ground reaction forces solely from a sparse set of six IMUs.}
      Importantly, PIP runs at $60$ frames per second with only $16\mathrm{ms}$ latency, which enables real-time applications.
     }
    \label{fig:teaser}
\end{figure}
\par
Capturing the motion of real humans is a long-standing and challenging problem with many applications in computer vision and graphics, movie production, gaming, AR, and VR.
However, due to its articulated structure, capturing the highly complex and potentially fast movements of the human body is challenging and many works have been proposed in the past~\cite{FullBodyZhou2021,Function4D,Chen2020,TransPose,EventCap,Wang2021}.
\par
One category of approaches are image-based where the actor motion is recovered by analyzing the image data, which can be either multi-view imagery~\cite{Chen2020,Zhang2020,Tome2018,Dong2019}, depth images~\cite{DoubleFusion,Function4D,DeepMultiCap,BodyFusion}, or a single RGB stream~\cite{XNect,DeepCap,FullBodyZhou2021,GraviCap,PyMAF,HMMR,Dwivedi2021,PARE}.
From the setting, it becomes clear that occlusions (either object-actor or self occlusions) can lead to a significantly reduced tracking quality.
Besides, these methods are sensitive to the lighting and the appearance of the actor as distinct features need to be extracted from images.
Moreover, many methods assume a static camera, resulting in a limited space where the subject can be captured.
These drawbacks limit the usability of optical motion capture.
\par
Recently, researchers start to explore alternative sensing devices such as inertial measurement units (IMUs).
Production-ready solutions~\cite{Xsens,Noitom} can track the body motion accurately \RED{solely from inertial sensors}.
However, they rely on special suits with densely placed sensors (usually 17 IMUs), which are difficult to wear.
Besides, the large number of IMUs can hinder the actor's movement.
Having a sparser set of IMUs on the body is clearly advantageous and more flexible.
However, recent sparse methods~\cite{SIP,DIP,TransPose} struggle with physical correctness and cannot disambiguate poses with similar sensor measurements such as sitting and standing; they are non-causal, \textit{i.e.}, need future information, which introduces large delays; their accuracy is still limited while temporal artifacts such as jitter become visible.
\par
To this end, we propose Physical Inertial Poser (PIP), a new real-time method for motion capture as well as joint torque and ground reaction force estimation using only six IMUs (see Fig.~\ref{fig:teaser}).
In contrast to previous works~\cite{DIP, TransPose} that require future information, our method only requires the information already available at any given time, which means no additional delay is introduced.
Our algorithm has two stages: \textit{1)} learning-based motion estimation and \textit{2)} physics-based motion optimization, which leverage both human kinematics and dynamics in motion capture.
\par
In the estimation stage, we regress human pose, joint velocities, and foot-ground contact probabilities from the inertia inputs using recurrent neural networks (RNNs).
We estimate leaf-to-full joint positions as intermediate tasks to improve the tracking accuracy as proposed by TransPose~\cite{TransPose}.
To resolve the pose ambiguity arising from the sparse IMU placement, we further propose a \textit{learning-based} RNN state initialization strategy, which helps the networks better learn the change of body pose from input inertia measurements.
This results in a significant accuracy improvement especially for ambiguous motions such as sitting still.
\par
In the optimization stage, we recover the physically correct motion, joint torques, and ground reaction forces from the kinematic estimations, leveraging a torque-controlled floating-base simulated character model.
Different from previous works that independently control the rotation of each degree of freedom of the character using proportional-derivative (PD) rules~\cite{PhysCap,PhysAware,Isogawa2020,SimPoE}, we propose a novel \textit{dual PD controller} to incorporate the global holistic control of the character's pose.
This is achieved by applying PD rules on both joint positions and rotations.
The proposed technique significantly improves the translation accuracy and physical plausibility of the motion.
\par
In summary, our main contributions are:
\setlist{nolistsep}
\begin{itemize}[noitemsep]\itemsep0em
    \item The first physics-aware real-time approach that estimates human motion, joint torques, and ground reaction forces with only six IMUs, which we call PIP.
    \item A learning-based RNN state initialization scheme, which helps to better disambiguate human motion regression from sparse IMU measurements (Sec.~\ref{sec:method-kinematics}).
    \item A dual PD controller, which achieves the combined control of local and global pose to improve the motion tracking accuracy and physical plausibility (Sec.~\ref{sec:method-dynamics}).
\end{itemize}
Our experiments demonstrate that PIP significantly outperforms previous sparse IMU-based methods in terms of tracking accuracy, physical plausibility, and disambiguation of challenging poses.
\section{Related Work}
Human motion capture (mocap) has a long research history.
Many works have been devoted to this topic, which can be mainly categorized into optical, inertial, and hybrid approaches.
Since our method only requires IMU measurements as input, we do not discuss purely image-based approaches~\cite{LiveCap,VNect,VIBE,HMRkanazawa2018,ROMP,PIFuHD,DeepCap}.
Here, we focus on hybrid and inertial mocap solutions, and the previous efforts on the physical plausibility of human motion.
\par\noindent\textbf{Optical-inertial Hybrid Motion Capture.}
As image-based mocap solutions suffer from occlusions, fusing images with IMUs, which aims at achieving more robust motion tracking, has recently attracted much attention.
This can be achieved by either energy-based optimization~\cite{Pons2010,Malleson2017,VIP,Marcard2016,Malleson2020,Kaichi2020} which optimizes human pose to fit both image features and inertia measurements, or feature-based estimation~\cite{Gilbert2019,TotalCapture} which regresses human pose from the combined features derived from images and IMUs.
Zhang et al.~\cite{FusingIMUImagesZhang2020} propose to exploit IMUs in the 2D pose estimation by fusing the image features of each pair of joints linked by the IMUs.
Some works fuse IMUs with depth images~\cite{Kalkbrenner2014,Zheng2018HybridFusion,Helten2013} or optical markers~\cite{Andrews2016} to perform human motion/performance capture.
Nevertheless, these methods are still substantially limited under low light conditions and heavy occlusions, and require the actor to move within the viewing frustum of the camera.
Our method requires no visual input, and thus is free from these limitations.
\par\noindent\textbf{Motion Capture from Inertial Sensors.}
Inertial mocap approaches do not suffer from occlusions or restricted moving space.
Commercial solutions~\cite{Xsens,Noitom} and the extended work~\cite{HPS} rely on 17 IMUs to perform motion capture.
They usually require the actor to wear a tight suit with densely bounded IMUs, which is inconvenient, intrusive, and obstructive.
It is clear that having a reduced set of IMUs on the body is preferable.
However, motion capture from sparse inertial sensors is very ambiguous and challenging.
Some works~\cite{Vlasic2007,Liu2011,Sy2020} leverage ultrasonic sensors for additional position information to resolve some ambiguities, but the use of distance sensors limits the recording range.
Early purely-inertial works~\cite{Slyper2008,Tautges2011,Riaz2015} use sparse accelerometers to reconstruct human pose by database search.
Schwarz et al.~\cite{Schwarz2009} use sparse orientation measurements to perform person-specific pose estimation.
To improve the accuracy, recent works~\cite{SIP,DIP,TransPose,Geissinger2020,Patrik2021} leverage both acceleration and orientation measurements.
Marcard et al.~\cite{SIP} present an offline method for human motion capture from only 6 IMUs, which achieves promising accuracy.
Huang et al.~\cite{DIP} propose the first deep learning method, which uses a bidirectional recurrent neural network (biRNN) to estimate the human pose from 6 IMUs in real-time.
However, their method does not allow to locate the person in the 3D space, \textit{i.e.,} the root translation is not estimated.
The current state-of-the-art method, TransPose~\cite{TransPose}, introduces the first real-time pose and translation estimation framework, which achieves an accurate capture quality while also only using 6 IMUs.
However, all of these works have a non-negligible delay due to the inherent need of future information, cannot stably capture ambiguous poses, and has many non-physical artifacts such as jitter and foot-sliding.
In contrast, our method does not rely on any future information and is free from the delay while even achieving higher accuracy.
In addition, we are the first to combine physics-based motion optimization with sparse inertial motion capture, and we show that such a carefully orchestrated design significantly improves the physical correctness of the motion.
\par\noindent\textbf{Physical Plausibility of Motion.}
To ensure the physical plausibility of motion, many works address the awareness of physics in their approach.
One category of works only impose physical constraints (\textit{e.g.,} foot contacts~\cite{MotioNet,Zou2020,Du2019}, temporal consistency~\cite{VNect,SIP,LiveCap}, and collision~\cite{Zanfir2018}) without considering human dynamics such as forces and masses.
Some works leverage an explicitly reconstructed scene to constrain the motion \cite{Hassan2019,LEMO,HPS}.
However, due to the articulated structure of humans, it is considerably difficult to track the complicated body movements by imposing such naive constraints.
Another category of works leverage physics-based human models and estimate forces to control the motion, targeting a more accurate modeling of real-world human movements.
Some works~\cite{PhysCap,Wei2010,Vondrak2012,Zell2017,Li2019,Rempe2020} use optimization-based methods to solve the optimal forces and human motion, which satisfy the physical constraints and laws such as the equation of motion~\cite{RigidBodyDynamics}.
Zell et al.~\cite{Zell2020} propose a weakly-supervised learning framework for dynamics estimation from human motion.
Shimada et al.~\cite{PhysAware} present a fully-differentiable framework for learning-based motion and force estimation from videos.
Reinforcement learning is also used in physics-based character control~\cite{Liu2018,DeepMimic,SFV,Yuan2019,Bergamin2019,Isogawa2020,SimPoE,Yu2021}, which can utilize advanced non-differentiable physics simulators.
Among these works, our physics module is most similar to the work of Shimada et al.~\cite{PhysCap}.
The major differences are the input to the respective method and the control of the physical character.
Our method assumes sparse inertia measurements of the moving body as input while theirs~\cite{PhysCap} leverages images of the actor.
Moreover, their approach~\cite{PhysCap} uses a proportional-derivative (PD) controller to control the rotation of each joint of the character independently.
In contrast, our method uses a novel \textit{dual PD controller} to introduce the global control of the character, aiming at better accuracy.
In other words, our method is the first that leverages explicit physics-based optimization into spare IMU-based motion capture.

\section{Method}
\label{sec:method}
\begin{figure*}
    \includegraphics[width=\textwidth]{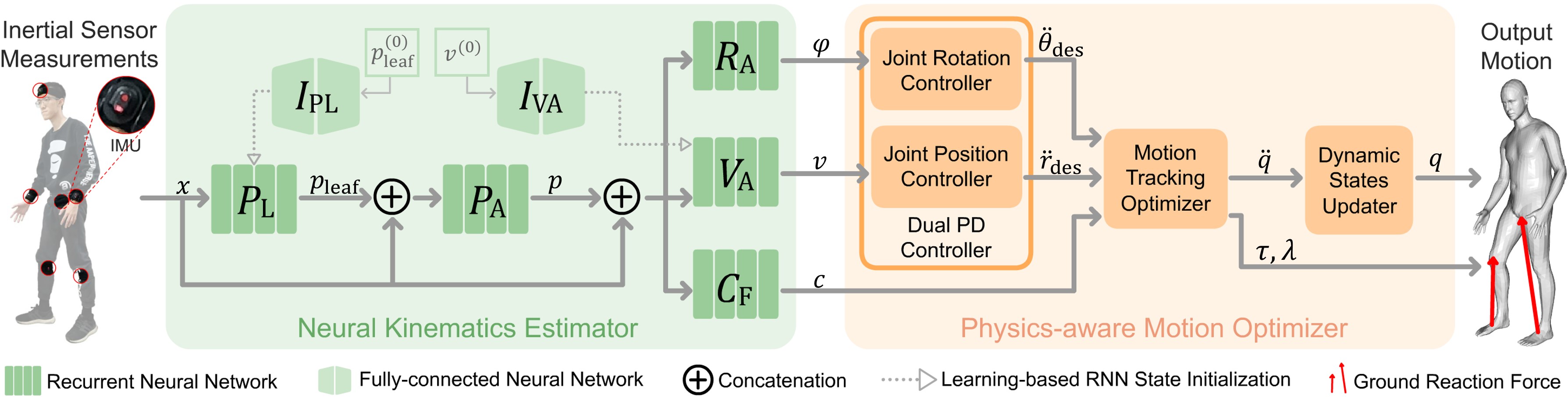}
    \caption{
        Overview of our method.
        We first use a neural kinematics estimator to infer human motion status from sparse IMU measurements.
        Then, we use a physics-aware motion optimizer to obtain physically correct human motion, joint torques, and ground reaction forces.
    }
    \label{fig:pipeline}
\end{figure*}
Our task is to track human motion in real-time using 6 IMUs.
The input of our method is the sequential measurements of accelerations and orientations of the 6 IMUs mounted on the left/right forearms, left/right lower legs, head, and pelvis (Fig.~\ref{fig:pipeline}).
The output of our method is the subject's motion in terms of joint angles and global translation, together with physical properties including ground reaction forces and joint torques.
The method incorporates two modules:
\textit{1)} \textit{the kinematics module}: a neural kinematics estimator, which infers the human motion from the IMU measurements, followed by
\textit{2)} \textit{the dynamics module}: a physics-aware motion optimizer, which refines the human motion and outputs the physical properties.

\subsection{Neural Kinematics Estimator}
\label{sec:method-kinematics}
The task of the kinematics module is to estimate the current motion status (specified later).
We use the same kinematic tree as in SMPL~\cite{SMPL}, which contains $J=24$ joints.
We refer to the wrists, ankles, and head as the \textit{leaf joints}, and the pelvis as the \textit{root joint}.
For all the $J$ joints, their 3D positions are denoted as $\boldsymbol p \in \mathbb{R}^{3J}$; their linear velocities are denoted as $\boldsymbol v \in \mathbb{R}^{3J}$; and their rotations are denoted as $\boldsymbol \varphi \in \mathbb{R}^{6J}$ in the 6D representation~\cite{6D}.
Since the IMUs do not provide any positional measurement, all these estimations are in the local coordinate frame (relative to the root joint).
Similar to TransPose~\cite{TransPose}, we perform a T-Pose calibration at the beginning, and then at each time step we stack the IMU measurements, \textit{i.e.,} the aligned accelerations and rotation matrices, into a single input vector $\boldsymbol x \in \mathbb{R}^{72}$.
In the following, we first give an overview of our network structure (Sec.~\ref{sec:method-kinematics-framework}).
Then, we dive into our novel learning-based initialization for the RNN hidden states during training (Sec.~\ref{sec:method-kinematics-RNN}).
The new initialization method helps the network learn to capture the state-change signals, which is crucial to resolve the pose ambiguity in our task.
%
%
\subsubsection{Motion Estimation Network}\label{sec:method-kinematics-framework}
The structure of the network follows the one of TransPose~\cite{TransPose}, which first estimates leaf joint properties and then full body status in a multi-stage style.
Different from their method~\cite{TransPose}, we choose to use RNN instead of biRNN as the basic network structure (elaborated in Sec.~\ref{sec:method-kinematics-RNN}), and we estimate the velocity of \textit{all joints} instead of \textit{only the root joint}, since we find that using full-joint velocities in combination with the physics part allows better character control.
\par
Specifically, as shown in Fig.~\ref{fig:pipeline}, we first use an RNN $P_\mathrm{L}$ to regress leaf joint positions $\boldsymbol p_{\mathrm{leaf}} \in \mathbb{R}^{15}$ from the IMU measurements $\boldsymbol x$.
Then, the concatenated vector $[\boldsymbol p_{\mathrm{leaf}}\,\,\,\,\boldsymbol x]$ is fed into the second RNN $P_\mathrm{A}$, which estimates all joint positions $\boldsymbol p \in \mathbb{R}^{3J}$.
Next, we feed the vector $[\boldsymbol p\,\,\,\,\boldsymbol x]$ into three RNNs $R_\mathrm{A}$, $V_\mathrm{A}$, and $C_\mathrm{F}$ to estimate the joint rotations $\boldsymbol \varphi \in \mathbb{R}^{6J}$ (the root orientation is directly measured by the IMU placed on the pelvis), linear velocities $\boldsymbol v \in \mathbb{R}^{3J}$, and foot-ground contact probabilities $\boldsymbol c \in \mathbb{R}^2$.
Finally, $\boldsymbol \varphi$, $\boldsymbol v$, and $\boldsymbol c$, which we call \textit{motion status}, are fed into the subsequent dynamics module.
During training, we use an L2 loss for $P_\mathrm{L}$, $P_\mathrm{A}$, and $R_\mathrm{A}$, a binary cross-entropy loss for $C_\mathrm{F}$, and the cumulative loss proposed in~\cite{TransPose} for $V_\mathrm{A}$.
%
%
\subsubsection{RNN with Learning-based Initialization}\label{sec:method-kinematics-RNN}
Full-body motion tracking from sparse inertial sensors is severely under-constrained and ambiguous.
For example, due to the sparsity of the IMUs, it is impossible to distinguish \textit{standing still} and \textit{sitting still} since the IMU measurements are identical: orientations are the same and accelerations are zero.
To cope with this ambiguity, leveraging the temporal information by capturing and memorizing the state-change signals in historical frames is a must.
Previous state-of-the-art works~\cite{DIP,TransPose} leverage bidirectional recurrent neural networks (biRNN)~\cite{biRNN} to learn such temporal information.
However, the design of biRNNs only allows a fixed frame window in the real-time setting, which prevents the access to state changes that happened outside this temporal window, \textit{e.g.,} when the subject remains seated for a longer time.
In consequence, these methods fail to capture such ambiguous poses correctly.
To overcome this limitation, we use RNNs to retain complete historical information and capture the crucial state-change signals.
\par
To capture the state-change signals, not only does the architecture have to be updated, but also a new training strategy is necessary.
Traditionally, an RNN is trained in a mini-batch manner and always starts with a zero initialization for hidden states in each batch.
However, in our setting, a constant initial state is incorrect (the subject may start from sitting, standing, lying, \textit{etc.});
and when the initial state is wrong, the model can never learn how to change its hidden state according to the signals afterward due to the mismatch at the beginning.
To address this problem, we propose a learning-based RNN initialization strategy.
Specifically, we have a separate fully-connected neural network (FCN), which regresses the initial state of the RNN from body pose information.
The FCN and RNN are trained jointly: for each mini-batch, the ground-truth pose at the beginning is fed into the FCN, then the output of the FCN is assigned to the hidden state of the RNN, then the RNN is trained as usual.
As the RNN implementation is not modified, the proposed strategy is highly effective and compatible with highly optimized black-box RNN libraries.
During inference, we assume the initial pose of the subject is known, which can be obtained from the calibration step.
Notice that the FCN only initializes the RNN for the first frame.
The FCNs we use for initialization during training are shown in Fig.~\ref{fig:pipeline} as $I_\mathrm{PL}$ and $I_\mathrm{VA}$, which take the beginning leaf joint positions $\boldsymbol p_{\mathrm{leaf}}^{(0)}$ and joint velocities $\boldsymbol v^{(0)}$ as input, respectively.
This initialization is only applied to $P_\mathrm{L}$ and $V_\mathrm{A}$, which suffer most from the ambiguity.

\subsection{Physics-aware Motion Optimizer}
\label{sec:method-dynamics}
The output of the kinematics module may still contain artifacts like jitter and ground penetration.
We therefore introduce the dynamics module to explicitly apply the physical constraints as similar to~\cite{PhysCap}.
The input to this module is the motion status $\boldsymbol \varphi$, $\boldsymbol v$, and $\boldsymbol c$ estimated by the kinematics module, which serve as the \textit{reference} in the physics-based optimization.
The task of the dynamics module is to obtain the motion, internal joint torques, and ground reaction forces that align with the reference but also satisfy physical constraints.
Specifically, based on the physics model (Sec.~\ref{sec:method-dynamics-model}), we first use a novel \textit{dual PD controller} (Sec.~\ref{sec:method-dynamics-controller}) to compute the desired acceleration for the simulated character which can fully reproduce the reference motion, and then use a motion optimizer (Sec.~\ref{sec:method-dynamics-optimizer}) to solve for the acceleration and force that the character can actually produce within the physical constraints.
Finally, we update the character status and compute the final output motion (Sec.~\ref{sec:method-dynamics-updater}).
%
%
\subsubsection{Physics Model}\label{sec:method-dynamics-model}
We use a torque-controlled floating-base simulated character~\cite{Zheng2013} as our physics model and follow the same mass distribution as in \cite{PhysCap}.
We initialize the subject's global position at the origin.
We refer to the joint positions in the global coordinate frame as $\boldsymbol r \in \mathbb{R}^{3J}$ and the translation as $\boldsymbol r_{\mathrm{root}} \in \mathbb{R}^3$.
The time derivative $\dot{\boldsymbol r}$ and $\ddot{\boldsymbol r}$ refer to the linear velocity and acceleration in the global frame.
We refer to the local joint rotations (\textit{i.e.,} pose) in Euler angles as $\boldsymbol \theta \in \mathbb{R}^{3J}$, and its time derivative $\dot{\boldsymbol \theta}$ and $\ddot{\boldsymbol \theta}$ are the angular velocity and acceleration.
The configuration of the character is described by its pose and translation, which we denote as $\boldsymbol q=[\boldsymbol r_{\mathrm{root}}\,\,\,\,\boldsymbol \theta] \in \mathbb{R}^N$ where $N=3+3J$ is the degree of freedom (DoF).
The time derivative $\dot{\boldsymbol q}$ and $\ddot{\boldsymbol q}$ are the generalized velocity and acceleration.
The character is controlled by the vector of force $\boldsymbol \tau \in \mathbb{R}^N$ where each dimension refers to the force on the corresponding DoF.
In the real world, the character is actuated only by the torques at the non-root joints, while no force is applied to the root joint.
However, to compensate for the dynamics mismatch between our physics model and real humans, we allow a small residual force at the root joint as prior works~\cite{PhysCap,ResidualForceControl,PhysAware,SimPoE,PhysSimulationLevine2012} do.
In our notation, the first six entries $\boldsymbol \tau_{:6}$ correspond to the residual force at the root joint, and $\boldsymbol \tau_{6:}$ are the actuated joint torques.
The generalized acceleration $\ddot{\boldsymbol q}$ and the force $\boldsymbol \tau$ follow the equation of motion~\cite{RigidBodyDynamics}:
\begin{equation}
    \boldsymbol \tau + \boldsymbol J_\mathrm{c}(\boldsymbol q)^T\boldsymbol \lambda = \boldsymbol M(\boldsymbol q)\ddot{\boldsymbol q} + \boldsymbol h(\boldsymbol q, \dot{\boldsymbol q}),
\end{equation}
where $\boldsymbol M \in \mathbb{R}^{N \times N}$ is the inertia matrix;
$\boldsymbol h \in \mathbb{R}^N$ is the non-linear effect term that accounts for gravity, Coriolis, and centripetal forces;
$\boldsymbol \lambda \in \mathbb{R}^{3n_\mathrm{c}}$ is the external contact forces applied at $n_\mathrm{c}$ character-ground contact points;
$\boldsymbol J_\mathrm{c} \in \mathbb{R}^{3n_\mathrm{c} \times N}$ is the contact point Jacobian, which maps the generalized velocity $\dot{\boldsymbol q}$ to the contact point velocities:
\begin{equation} \label{eq:Jqdot}
    \dot{\boldsymbol r}_\mathrm{c} = \boldsymbol J_\mathrm{c} \dot{\boldsymbol q}.
\end{equation}
Readers are referred to~\cite{RigidBodyDynamics} for more details.
In our model, we assume all external forces \RED{(except for gravity)} are the \RED{support and frictional forces exerted at the contact points by the ground}, which we call \textit{ground reaction force (GRF)}.
%
%
\subsubsection{Dual Proportional-differential Controller}\label{sec:method-dynamics-controller}
To control the character, previous methods~\cite{PhysCap,PhysAware,Isogawa2020,SimPoE} use a single PD controller to compute either angular accelerations or joint torques to reproduce the reference motion.
However, since the configuration of the character is parameterized in \textit{local} Euler angles, such methods only focus on the independent control of local joint rotations, which may result in an undesirable global pose.
Simply applying PD control on \textit{global} joint rotations will make the optimization problem non-quadratic, introducing a large computation cost.
We find that imposing a PD controller on joint positions will constrain the global pose, while still keeping the problem quadratic.
To this end, we propose a \textit{dual PD controller}, which contains \textit{1)} a rotation controller controlling the local pose in joint rotational space and \textit{2)} an additional position controller controlling the global pose in joint positional space.
\RED{Below, we elaborate the two controllers.}
\par\noindent\textbf{Joint Rotation Controller.}
This controller computes the desired joint angular acceleration $\ddot{\boldsymbol \theta}_\mathrm{des}$ from the estimated reference joint rotations $\boldsymbol \varphi$ using:
\begin{equation}
    \ddot{\boldsymbol \theta}_\mathrm{des} = k_{p_\theta}(\mathbf{E}(\boldsymbol \varphi) - \boldsymbol \theta) - k_{d_\theta}\dot{\boldsymbol \theta},
\end{equation}
where $\boldsymbol \theta$ and $\dot{\boldsymbol \theta}$ are the current joint angles and angular velocities; $\mathbf E(\cdot)$ transforms the reference pose to local Euler angles; $k_{p_\theta} = 2400$ and $k_{d_\theta} = 60$ are the gain parameters.
\par\noindent\textbf{Joint Position Controller.}
This controller computes the desired linear joint acceleration $\ddot{\boldsymbol r}_\mathrm{des}$.
Different from the rotation controller, we do not have reference joint positions since we have no direct distance measurements.
Thus, we compute them from the current joint positions $\boldsymbol r$ and the estimated joint velocity $\boldsymbol v$ as:
\begin{equation}
    \boldsymbol r_{\mathrm{ref}} = \boldsymbol r + \mathbf T(\boldsymbol v)\Delta t,
\end{equation}
where $\mathbf T(\cdot)$ maps the joint velocity from local frame to global frame and $\Delta t$ is the simulation time step.
Then, the joint position controller is defined as:
\begin{equation}
    \ddot{\boldsymbol r}_\mathrm{des} = k_{p_r}(\boldsymbol r_{\mathrm{ref}} - \boldsymbol r) - k_{d_r}\dot{\boldsymbol r},
\end{equation}
where $k_{p_r} = 3600$ and $k_{d_r} = 60$ are the gain parameters; $\dot{\boldsymbol r}$ is the current joint velocity, which can be computed by:
\begin{equation}
    \dot{\boldsymbol r} = \boldsymbol J \dot{\boldsymbol q},
\end{equation}
where $\boldsymbol J \in \mathbb{R}^{3J\times N}$ is the joint Jacobian.
%
%
%
\subsubsection{Motion Tracking Optimizer}\label{sec:method-dynamics-optimizer}
The motion tracking optimizer solves a quadratic programming problem and estimates acceleration $\ddot{\boldsymbol q}$, joint torques $\boldsymbol \tau$, and GRF $\boldsymbol \lambda$.
The optimization problem can be written as:
\begin{equation}\label{eq:qp}
\begin{array}{rll}
    \mathop{\arg\min}\limits_{\ddot{\boldsymbol q},\boldsymbol \lambda,\boldsymbol \tau}&\mathcal{E}_{\mathrm{PD}} + \mathcal{E}_{\mathrm{reg}}&\\
    \mathrm{s.t.}&\boldsymbol \tau + \boldsymbol J_\mathrm{c}^T\boldsymbol \lambda = \boldsymbol M\ddot{\boldsymbol q} + \boldsymbol h&(\mathrm{equation\,\,of\,\,motion})\\
    &\boldsymbol\lambda \in \mathcal{F}&(\mathrm{friction\,\,cone})\\
    &\dot{\boldsymbol r}_\mathrm{j}(\ddot{\boldsymbol q}) \in \mathcal{C}&\mathrm{(no\,\,sliding)}.
\end{array}
\end{equation}
$\boldsymbol M$ is computed from $\boldsymbol q$ using the composite rigid body algorithm~\cite{RigidBodyDynamics}.
$\boldsymbol h$ is computed from $\boldsymbol q$ and $\dot{\boldsymbol q}$ using the recursive Newton-Euler algorithm~\cite{RigidBodyDynamics}.
The energy function and the three constraints will be elaborated in the following.
\par\noindent\textbf{Contact Point Determination.}
To apply the three constraints in Eq.~\ref{eq:qp}, we first need to acquire all the contact points between the body and the ground.
We determine whether joint $j$ contacts the ground by its vertical distance to the ground $d_j$, and for the foot joint we additionally leverage the predicted contact probability $c$ for better accuracy since feet touch the ground more often.
Specifically, a foot joint $f$ is considered in contact if \textit{1)} $d_f < 0.5\mathrm{cm}$ or \textit{2)} $d_f < 3\mathrm{cm}$ and \RED{the ground contact probability} $c_f>0.5$; a non-foot joint $n$ is considered in contact only if $d_n < 0.5\mathrm{cm}$.
We then draw an $L \times L$ square centered at each contact joint and take its 4 vertices as the contact points.
This is based on our finding that assuming facet-contacts instead of point-contacts produces more stable results.
We take $L=20\mathrm{cm}$ which is roughly the size of a foot.
The number of the contact joints is denoted as $n_\mathrm{j}$, then the number of the contact points is $n_\mathrm{c}=4n_\mathrm{j}$.
\par\noindent\textbf{Dual PD Controller Term} $\boldsymbol{\mathcal{E}_{\mathrm{PD}}}$\textbf{.}
To reproduce the kinematic estimation, the character should generate the angular and linear joint accelerations given by the dual PD controller.
Thus, in Eq.~\ref{eq:qp}, $\mathcal{E}_{\mathrm{PD}}$ consists of two components $\mathcal{E}_\theta$ and $\mathcal{E}_r$, which control the angular and linear accelerations:
\begin{equation}
\begin{aligned}
    &\mathcal{E}_{\mathrm{PD}} = k_\theta\mathcal{E}_\theta + k_r\mathcal{E}_r,\\
    &\mathcal{E}_\theta = \|\ddot{\boldsymbol q}_{3:} - \ddot{\boldsymbol \theta}_\mathrm{des}\|^2,
    \mathcal{E}_r = \|\boldsymbol J\ddot{\boldsymbol q} + \dot{\boldsymbol J}\dot{\boldsymbol q} - \ddot{\boldsymbol r}_\mathrm{des}\|^2,\\
\end{aligned}
\end{equation}
where $k_\theta$ and $k_r$ are the weight terms both set to $1$.
\par\noindent\textbf{Regularization Term} $\boldsymbol{\mathcal{E}_{\mathrm{reg}}}$\textbf{.}
Our regularization term $\mathcal{E}_\mathrm{reg}$ in Eq.~\ref{eq:qp} contains three energy terms: \textit{1)} $\mathcal{E}_\lambda$ penalizes violations of the Signorini's conditions of contacts~\cite{signorini}; \textit{2)} $\mathcal{E}_\mathrm{res}$ constrains the magnitude of the root residual force; and \textit{3)} $\mathcal{E}_\tau$ confines the norms of the joint torques:
\begin{equation}
\begin{aligned}
    &\mathcal{E}_{\mathrm{reg}} = k_\lambda\mathcal{E}_\lambda + k_\mathrm{res}\mathcal{E}_\mathrm{res} + k_\tau\mathcal{E}_\tau,\\
    &\mathcal{E}_\lambda = \sum_{c=1}^{n_\mathrm{c}}d_c\|\boldsymbol \lambda_c\|^2,
    \mathcal{E}_\mathrm{res} = \|\boldsymbol \tau_{:6}\|^2,
    \mathcal{E}_\tau = \|\boldsymbol \tau_{6:}\|^2,\\
\end{aligned}
\end{equation}
where $d_c$ is the vertical height of the contact point $c$; $\boldsymbol \lambda_c$ is the GRF at point $c$; $k_\lambda$, $k_\mathrm{res}$, and $k_\tau$ are the corresponding weights, which are set to $10$, $0.1$, and $0.01$, respectively.
\par\noindent\textbf{Friction Cone and Sliding Constraints.}
These two constraints in Eq.~\ref{eq:qp} are only applied to the contacts.
We assume the GRF at the contact points should be inside the friction cone\footnote{\RED{Friction cone: the set of all forces that can be transmitted through a Coulomb friction contact. See~\cite{frictioncone}.}} and the contact joints do not slide.
Specifically, we denote the force/velocity along the $y$ (vertical) axis of the global frame as $\cdot^y$ and the same for the $x, z$ (horizontal) axis.
The friction cone constraint can be linearized as:
\begin{equation}
\begin{aligned}
    &\mathcal{F}_c=\{\boldsymbol\lambda_c\in\mathbb{R}^3|\boldsymbol\lambda_c^y \ge 0, |\boldsymbol\lambda_c^x| \le \mu \boldsymbol\lambda_c^y , |\boldsymbol\lambda_c^z| \le \mu \boldsymbol\lambda_c^y\},\\
    &\mathcal{F}=\{[\boldsymbol\lambda_1\cdots\boldsymbol\lambda_{n_\mathrm{c}}]\in\mathbb{R}^{3n_\mathrm{c}}|\boldsymbol\lambda_c\in\mathcal{F}_c,c=1,2,\cdots,n_\mathrm{c}\},
\end{aligned}
\end{equation}
which means the vertical force from the ground must be upward, and the horizontal forces should not be larger than the maximum frictional force.
We empirically set the friction coefficient $\mu = 0.6$.
For the sliding constraint, we have:
\begin{equation}
  \begin{aligned}
  &\mathcal{C}_j = \{\dot{\boldsymbol r}_j\in \mathbb{R}^3|\dot{\boldsymbol r}_j^y \ge 0,|\dot{\boldsymbol r}_j^x| \le \sigma, |\dot{\boldsymbol r}_j^z| \le \sigma\},\\
  &\mathcal{C} = \{[\dot{\boldsymbol r}_1\cdots\dot{\boldsymbol r}_{n_\mathrm{j}}]\in \mathbb{R}^{3n_\mathrm{j}}|\dot{\boldsymbol r}_j\in\mathcal{C}_j,j=1,2,\cdots,n_\mathrm{j}\},
  \end{aligned}
  \end{equation}
which confines the sliding velocity of every contact joint smaller than $\sigma = 0.01$ while preventing ground penetration.
The contact joint velocity $\dot{\boldsymbol r}_\mathrm{j}(\ddot{\boldsymbol q})$ in Eq. \ref{eq:qp} is computed by:
\begin{equation}
    \dot{\boldsymbol r}_\mathrm{j}(\ddot{\boldsymbol q})=\boldsymbol J_\mathrm{j}(\dot{\boldsymbol q} + \ddot{\boldsymbol q}\Delta t),\\
\end{equation}
where $\boldsymbol J_\mathrm{j} \in \mathbb{R}^{3n_\mathrm{j} \times N}$ is the contact joint Jacobian.
%
%
\subsubsection{Dynamic States Updater}\label{sec:method-dynamics-updater}
We use a finite difference method for dynamic state updates:
\begin{equation}
\begin{aligned}
    &\boldsymbol q^{(t+1)} = \boldsymbol q^{(t)} + \dot{\boldsymbol q}^{(t)}\Delta t,\\
    &\dot{\boldsymbol q}^{(t+1)} = \dot{\boldsymbol q}^{(t)} + \ddot{\boldsymbol q}^{(t)}\Delta t,\\
\end{aligned}
\end{equation}
where $\ddot{\boldsymbol q}^{(t)}$ is the estimated acceleration from the optimizer and $\boldsymbol q^{(t + 1)}$ is the updated pose and translation. 
\RED{Since our system runs at 60 fps,  $\Delta t$ is set to $1/60$ second.}

\section{Experiments}
\label{sec:experiments}
In this section, we first compare our approach with previous works (Sec.~\ref{sec:experiments-comparisons}).
Then, we perform an ablative study of the key components (Sec.~\ref{sec:experiments-evaluations}).
Finally, we show the potential applications of our methods (Sec.~\ref{sec:experiments-applications}).
%
\par\noindent\textbf{Datasets.}
The training and evaluation involve the AMASS dataset~\cite{AMASS}, the DIP-IMU dataset~\cite{DIP}, and the TotalCapture dataset~\cite{TotalCapture}.
Following~\cite{TransPose}, we first train the model on AMASS using synthesized IMUs and then fine-tune it on the train split of DIP-IMU.
The evaluations are performed on TotalCapture and the test split of DIP-IMU.
The acceleration measurement in TotalCapture is constantly biased and we re-calibrated it (\RED{detailed in the supplemental document}).
\RED{All the reported numbers are online results.}
%
%
\par\noindent\textbf{Metrics.}
We use the following metrics to evaluate our method.
\textit{1) SIP Error} measures the mean orientation error of the upper arms and legs \RED{in the global space} in degrees.
\textit{2) Mesh Error} measures the mean vertex distance between the reconstructed and ground-truth meshes \RED{with both root position and orientation aligned} in $\mathrm{cm}$.
\textit{3) Jitter} measures the mean jerk (time derivative of acceleration) of all body joints \RED{in the global space} in $\mathrm{km/s^3}$, which reflects the smoothness of the motion~\cite{Jerk}.
\textit{4) Zero-Moment Point (ZMP) distance} measures the mean distance from the fictitious ZMP~\cite{ZMP} position to the Base of Support\footnote{\RED{Base of Support (BoS): the area around the outside edge of the body sections in contact with the ground. Also called the support polygon.}} of the character in $\mathrm{m}$.
It represents the intensity of the perturbation moment which lets the character fall, and should be zero for a real human in dynamic equilibrium~\cite{ZMP}.
\RED{
Previous works~\cite{CoPStabilityECCV,CoPStabilityThesis} leverage Center of Pressure (CoP) to quantify the equilibrium, which is related to the ZMP distance.
A discussion about ZMP and CoP can be found in the supplemental document.}
Among these metrics, 1) and 2) measure pose accuracy, 3) and 4) measure physical plausibility.
We further evaluate \textit{cumulative translation error} which means the global position error \textit{w.r.t} the real travelled distance; and \textit{latency} \RED{which measures the time from receiving the inertia measurements to outputting the pose and translation for the corresponding frame} in $\mathrm{ms}$, using a laptop with an Intel(R) Core(TM) i7-10750H CPU and an NVIDIA RTX2080 Super graphics card.
For all these metrics, the lower, the better.

\subsection{Comparisons}
\label{sec:experiments-comparisons}
\begin{figure}
      \includegraphics[width=0.95\linewidth]{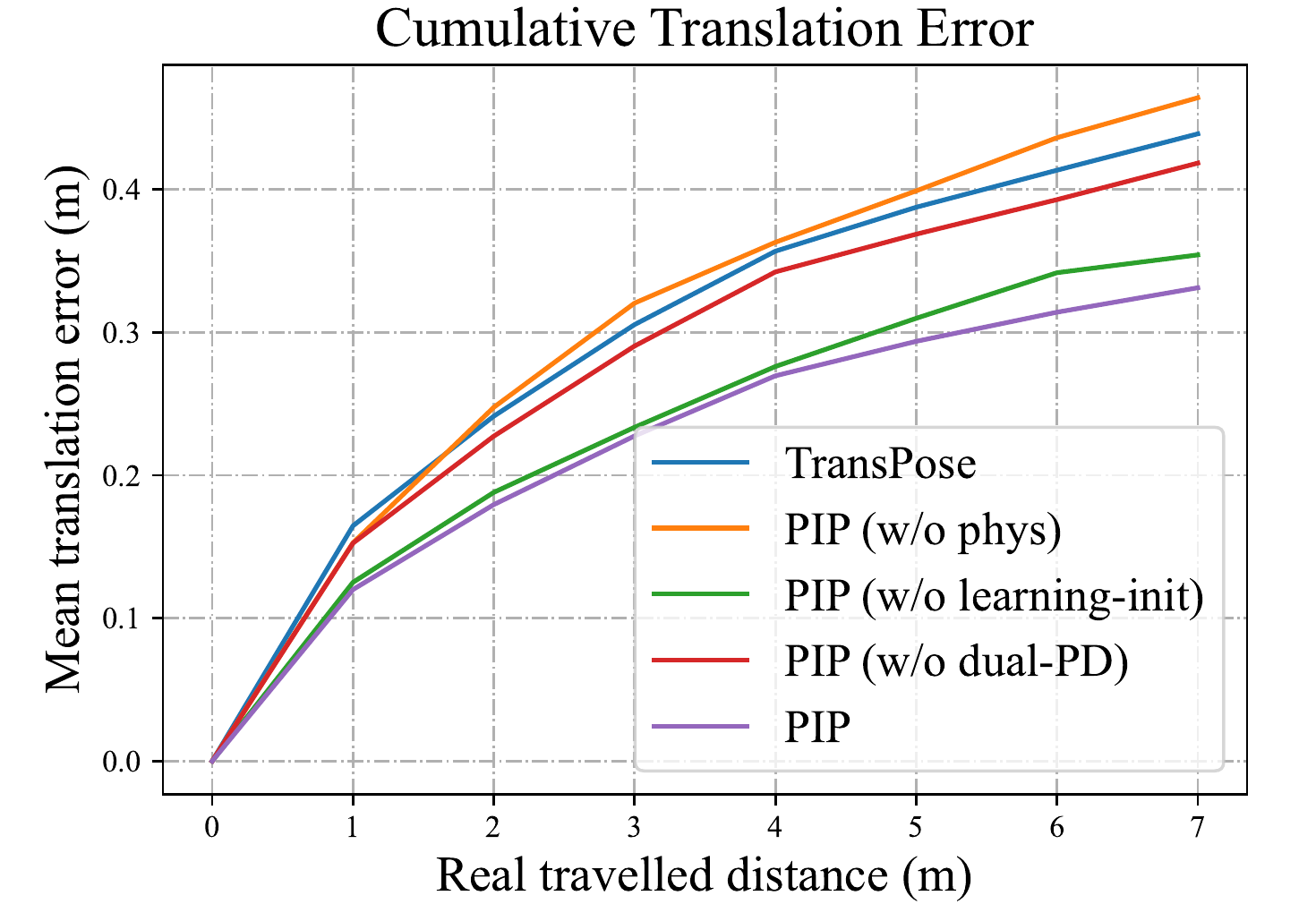}
      \centering
      \caption{
         Comparison of the translation estimation on TotalCapture~\cite{TotalCapture}.
         Our method has the lowest cumulative error because of the learning-based RNN initialization and the dual PD controller.
         %
      }
      \label{fig:tran_cmp}
  \end{figure}

\noindent\textbf{Quantitative.}
We compare our method to state-of-the-art methods DIP~\cite{DIP} and TransPose~\cite{TransPose} which also target motion capture from sparse IMUs.
Note that DIP does \textit{not} estimate global translation.
The results are shown in Tab.~\ref{tab:comparisons} and Fig.~\ref{fig:tran_cmp}.
Our method not only significantly outperforms previous works on capture accuracy and physical plausibility, but also largely reduces the delay.
The pose accuracy improvement is attributed to the RNN-based kinematics estimator which makes use of complete historical information and better captures state-change signals.
The improvement of the motion smoothness, equilibrium, and translation accuracy is attributed to the physics-aware motion optimizer with the novel dual PD controller.
Thus, the proposed combination of learning-based kinematics and optimization-based physics leads to the overall best result.
%
%
\begin{table}[t]
  \centering
  \resizebox{\linewidth}{!}{%
  \begin{tabular}{|c|c|c|c|c|c|}
  \hline
  \multirow{2}{*}{Method} & \multicolumn{5}{c|}{DIP-IMU}                                                     \\ \cline{2-6}
                              & SIP Err        & Mesh Err      & Jitter       & ZMP Dist      & Latency      \\ \hline
  DIP~\cite{DIP}              & 17.10          & 8.96          & -             & -             & 117          \\ \hline
  TransPose~\cite{TransPose}  & 16.68          & 7.09          & 1.46          & 1.67          & 94          \\ \hline
  \textbf{PIP}                & \textbf{15.02} & \textbf{5.95} & \textbf{0.24} & \textbf{0.12} & \textbf{16} \\ \hline
  \end{tabular}%
  }
  \vspace{-1.3em}
  \end{table}
  \begin{table}[t]
  \centering
  \resizebox{\linewidth}{!}{%
  \begin{tabular}{|c|c|c|c|c|c|}
  \hline
  \multirow{2}{*}{Method} & \multicolumn{5}{c|}{TotalCapture}                                               \\ \cline{2-6}
                              & SIP Err        & Mesh Err      & Jitter       & ZMP Dist      & Latency     \\ \hline
  DIP~\cite{DIP}              & 18.62          & 11.22         & -             & -             & 117         \\ \hline
  TransPose~\cite{TransPose}  & 16.58          & 7.42          & 1.87          & 1.40          & 94         \\ \hline
  \textbf{PIP }               & \textbf{12.93} & \textbf{6.51} & \textbf{0.20} & \textbf{0.23} & \textbf{16} \\ \hline
  \end{tabular}%
  }
  \caption{
  Comparison with the state-of-the-art methods on DIP-IMU~\cite{DIP} and TotalCapture~\cite{TotalCapture}.
  \RED{Metrics and units are detailed at the beginning of Sec.~\ref{sec:experiments}.
  PIP achieves a reduction of 15\% of the pose error, 87\% of the jitter, and 89\% of the motion imbalance with 83\% lower latency compared with the SOTA TransPose~\cite{TransPose}.}
  }
  \label{tab:comparisons}
  \end{table}
  %
  %
  %
  %
  \begin{figure}
      \includegraphics[width=\linewidth]{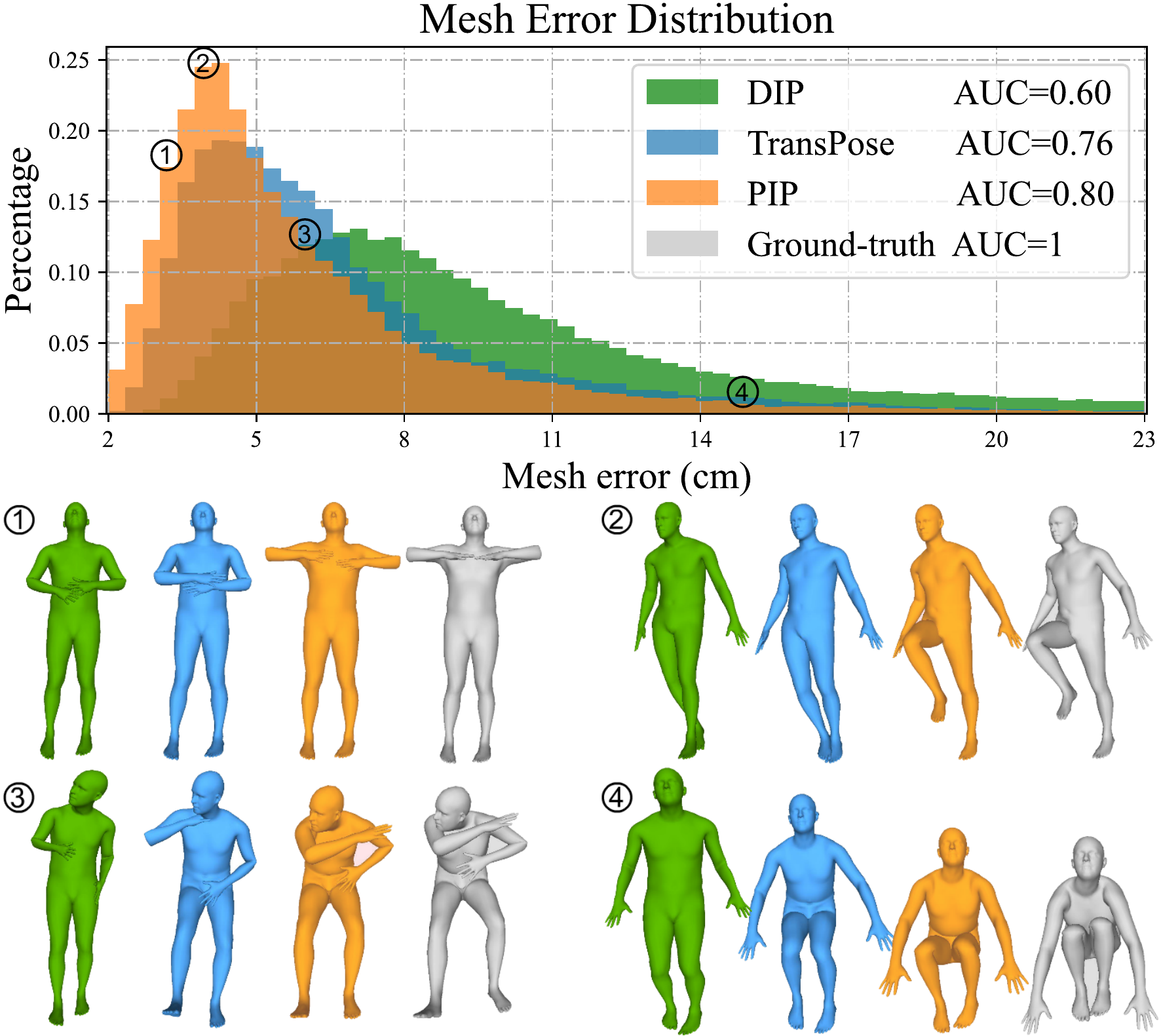}
      \caption{
          Qualitative results on TotalCapture \cite{TotalCapture} dataset.
          We show the mesh error distribution \RED{and the Area Under Curve (AUC) value} of different methods, and select 4 examples for visualization.
          %
          %
          Colors of different methods are shown in the legend.
          Our approach is visually the most accurate over all the methods.
      }
      \label{fig:qual_cmp}
  \end{figure}
  %
%
\par\noindent\textbf{Qualitative.}
In Fig.~\ref{fig:qual_cmp}, we show the mesh error distribution of DIP~\cite{DIP}, TransPose~\cite{TransPose}, and our method on TotalCapture.
We take 4 examples at \textit{1)} 10\%, \textit{2)} mode, \textit{3)} median, and \textit{4)} 95\%.
In the first two cases, our method estimates arm and leg orientations better than the previous works.
In the third case, we reconstruct the full-body pose faithfully while others nearly fail.
In the last challenging example, although the estimated upper legs slightly defer from the ground truth, our result still looks similar and outperforms others.
Again, we attribute this superiority to the RNN-based estimator and the physics-based optimizer.
The ambiguity in these cases comes from the fact that the subject can perform very different poses while keeping the forearm/lower leg orientation unchanged, and the key to resolving the ambiguity is the temporal information of the state-change signals.
Compared with previous works, we make better use of such information due to our learning-based RNN initialization.
In consequence, our networks regress more accurate pose and velocities, which, in combination with the dual PD controller, further improve the results.

\subsection{Evaluations}
\label{sec:experiments-evaluations}
%
\begin{table}[t]
\centering
\resizebox{\linewidth}{!}{%
\begin{tabular}{|c|c|c|c|c|}
\hline
\multirow{2}{*}{Method} & \multicolumn{2}{c|}{DIP-IMU}   & \multicolumn{2}{c|}{TotalCapture} \\ \cline{2-5}
                        & SIP Error        & Jitter        & SIP Error        & Jitter           \\ \hline
w/o learning-init       & 15.12          & 0.27          & 13.70          & 0.23             \\ \hline
w/o dual-PD             & 15.04          & 0.28          & 12.93          & 0.32             \\ \hline
w/o physics module             & 15.04          & 0.48          & \textbf{12.84} & 0.51             \\ \hline
Ours                    & \textbf{15.02} & \textbf{0.24} & 12.93          & \textbf{0.20}    \\ \hline
\end{tabular}%
}
\caption{
    Ablation study on the learning-based RNN initialization, the dual PD controller, and the physics-based optimizer.
    It demonstrates the help of our key components on pose accuracy (shown in SIP Error) and physical plausibility (shown in Jitter).
}
\label{tab:abl}
\end{table}
%
%
\begin{figure}
  \includegraphics[width=\linewidth]{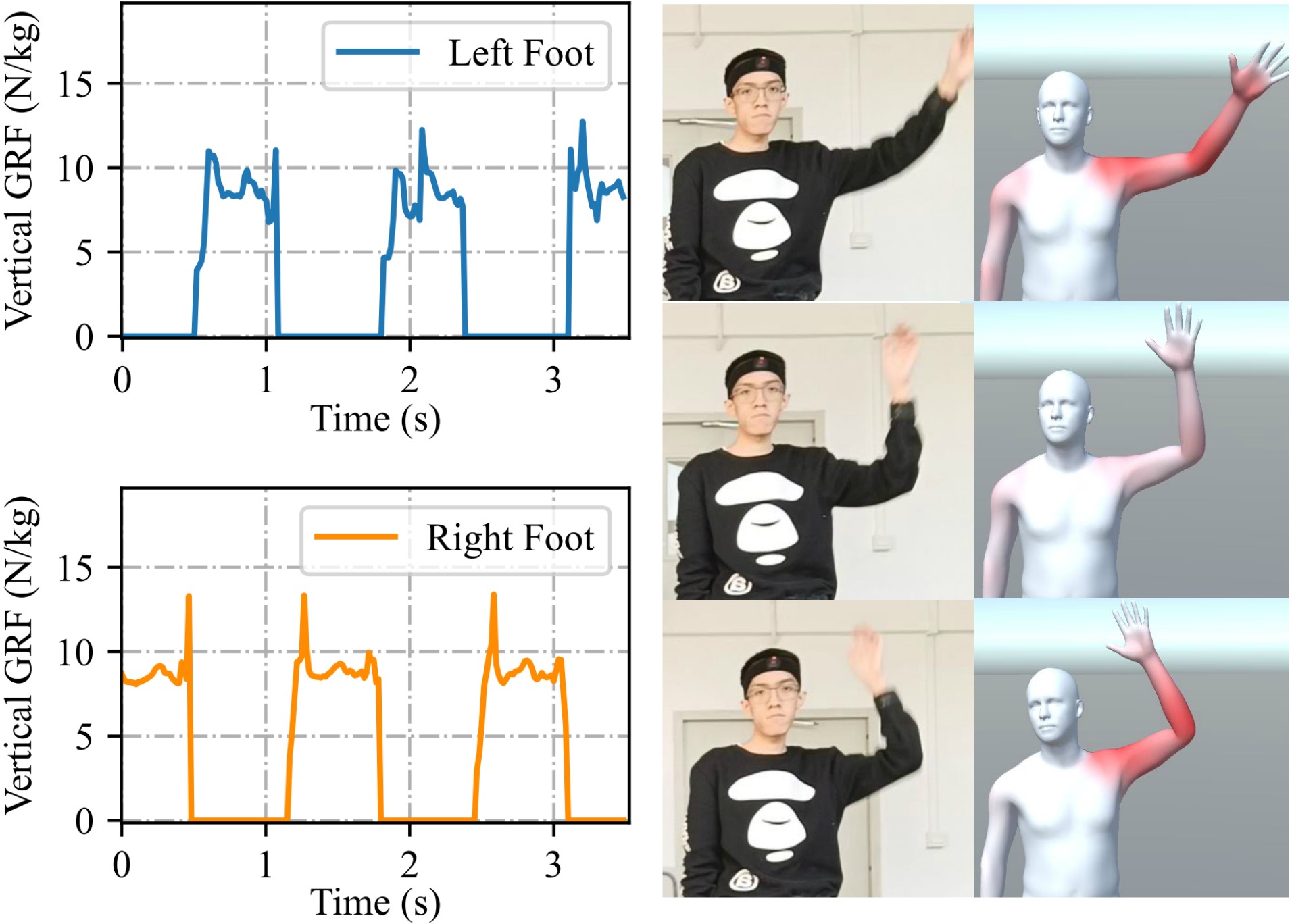}
  \caption{
      \textit{Left}: evaluation on the estimated contact forces over time in a walking motion.
      \textit{Right}: visualization of the arm joint torques for a waving-hand motion (red for large torques).
  }
  \label{fig:physeval}
\end{figure}
%
%
\noindent\textbf{Physical Properties.}
In Fig.~\ref{fig:physeval}, we demonstrate the estimated physical properties.
The left figure shows the GRF of two feet when the subject is walking.
We can see that the two feet support the body in turns and the GRF approximately equals gravity, which is reasonable according to~\cite{Shahabpoor2017,Zell2020}.
The right figure is a series of frames where the subject waves his hand and we visualize the internal torques of the arm (red for large torques).
When the arm starts and stops moving, the torque is larger due to the acceleration.
%
%
\par
\begin{figure}
    \includegraphics[width=\linewidth]{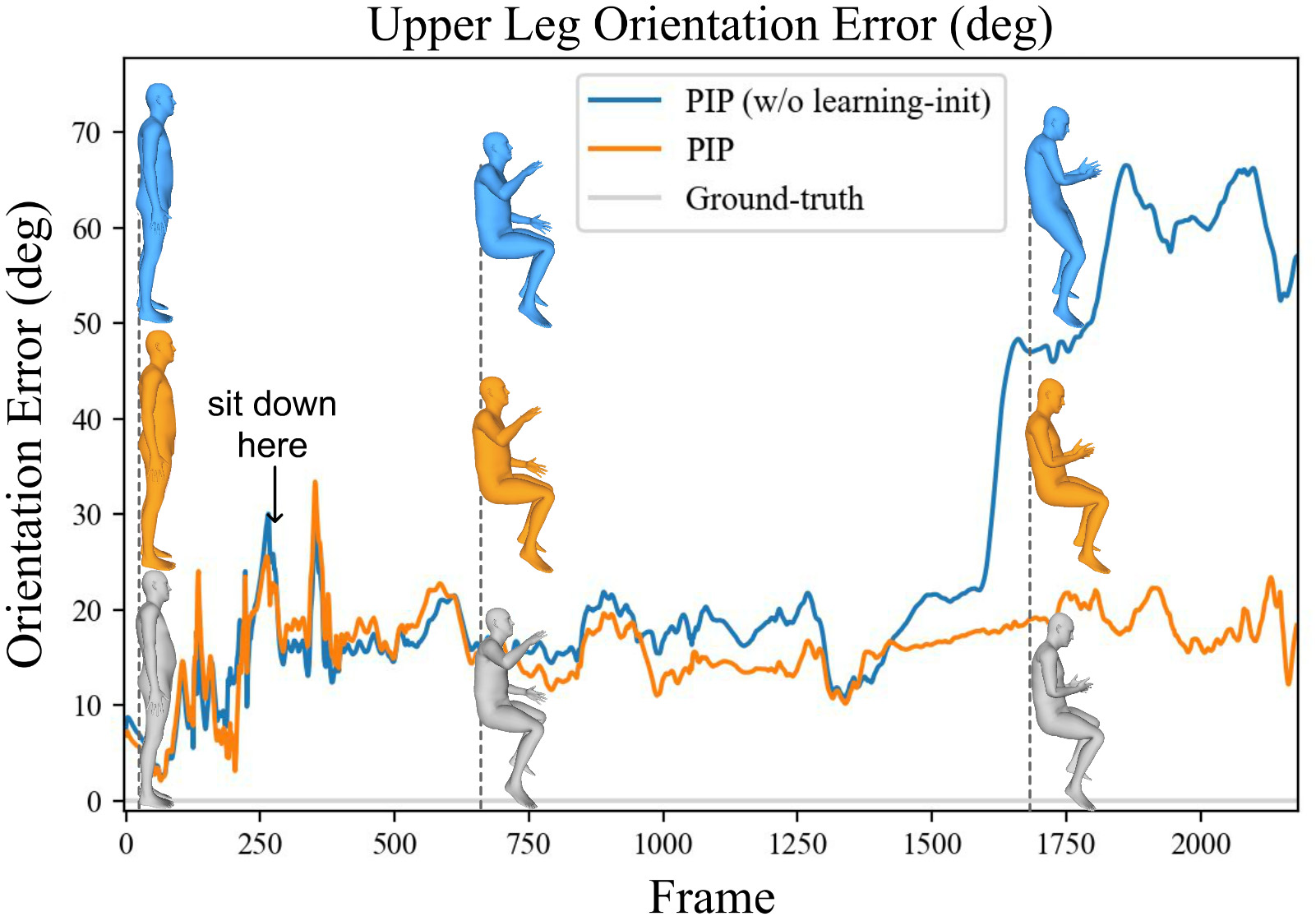}
    \caption{
       Ablation study on the learning-based RNN initialization skill using a long-sitting sequence.
       We plot the upper leg orientation error over time and pick three frames for visualization.
       %
    }
    \label{fig:abl}
\end{figure}

\noindent\textbf{Learning-based RNN Initialization.}
To examine the effect of the learning-based initialization, we train the same networks ($P_\mathrm{L}$ and $V_\mathrm{A}$) without learning-based initialization and perform the evaluation.
As shown in Tab.~\ref{tab:abl} and Fig.~\ref{fig:tran_cmp}, this variant becomes less accurate and stable, which is reflected in the larger errors across all metrics.
The learned initialization is most effective when the motion is highly ambiguous (\textit{e.g.,} sitting), but this advantage is numerically averaged out by the common (non-ambiguous) motion in the test dataset.
To this end, we pick a long-sitting sequence in the DIP-IMU test dataset and plot the upper leg orientation error over time in Fig.~\ref{fig:abl}.
During the sequence, the subject started from standing, then sat down immediately, and kept sitting to the end.
The curves show that the method without learning-based initialization starts to fail as time goes by while our model stably tracks the sequence.
%
%
By examining a few selected frames from the sequence, we can see that the zero-initialized version (in blue) is correct at the beginning but goes wrong after a long period because it loses the historical state-change information from standing to sitting.
As a result, its prediction stands up again.
In contrast, ours (in orange) always gives the correct estimation due to the good memorizing of such information.
%
%
%
%
\par\noindent\textbf{Physics and Dual PD Controller.}
We evaluate
\textit{1)} removing the physics-based optimizer, \textit{i.e.,} only using the kinematics module and integrating root velocities to obtain the translation;
and \textit{2)} removing the joint position controller, \textit{i.e.,} replacing the dual PD controller with a single PD controller that only watches $\boldsymbol q$ as in \cite{PhysCap}.
As shown in Fig.~\ref{fig:tran_cmp}, without the physics-based optimizer or the dual PD controller, the translation accuracy deteriorates significantly.
Besides, the method without physics-based optimization will \textit{always} estimate a character floating in the air or sinking into the ground due to the accumulated error of velocities, and the foot-sliding artifacts are also severe.
These facts demonstrate the necessity of physics awareness and our dual PD controller in mitigating translation error accumulation.
Quantitative results in Tab.~\ref{tab:abl} also show a significant reduction of the motion jitter in our full method.
However, with the physics-based optimization, the SIP error for TotalCapture is slightly larger ($0.1^{\circ}$).
This is because the physics module is mainly helpful for estimating translation and improving the physical correctness of the motion.

\subsection{Applications}
\label{sec:experiments-applications}
Our method enables several applications such as real-time animation of a virtual character and motion re-targeting.
Also note that reducing the latency from $94\mathrm{ms}$ to $16\mathrm{ms}$ is critical to enable applications such as gaming.
Please see our supplementary materials for more results.
%

\section{Conclusion and Limitations}
In this work, we present the first real-time physics-aware approach that estimates human motion, joint torques, and ground reaction forces from solely 6 IMUs.
Combining the kinematics and the physics modules leads to higher accuracy and realism, as shown in our experiments.
We also demonstrate exciting applications like live motion capture.
\par
%
%
However, we simplify the real world too much, \textit{e.g.,} assuming a flat ground, which makes our method incapable of capturing humans walking upstairs.
\RED{Besides, the current approach is based on the assumption of a known body shape.
For different body shapes, we only need to adjust the bone lengths and the mass distribution of the physics model.}

{\small
\bibliographystyle{ieee_fullname}
\bibliography{cv}
}

\appendix


\begin{table*}[t]
\centering
\resizebox{\linewidth}{!}{%
\begin{tabular}{|c|c|c|c|c|c|c|c|c|c|}
\hline
\multicolumn{2}{|c|}{\multirow{2}{*}{Method}}      & \multicolumn{8}{c|}{DIP-IMU}                                                                                                  \\ \cline{3-10} 
\multicolumn{2}{|c|}{\multirow{2}{*}{}}            & SIP Error      & Ang Error     & Pos Error     & Mesh Error    & Rel Jitter     & Abs Jitter    & ZMP Dist      & Latency     \\ \hline
\multirow{2}{*}{Offline} & DIP~\cite{DIP}          & 16.36          & 14.41         & 6.98          & 8.56          & 2.34           & -             & -             & -           \\ \cline{2-10} 
                         & TransPose~\cite{TransPose}& 13.97          & 7.62          & 4.90          & 5.83          & 0.13           & 0.85          & 0.59          & -           \\ \hhline{|==========|}
\multirow{3}{*}{Online}  & DIP~\cite{DIP}          & 17.10          & 15.16         & 7.33          & 8.96          & 3.01           & -             & -             & 117          \\ \cline{2-10} 
                         & TransPose~\cite{TransPose}& 16.68          & 8.85          & 5.95          & 7.09          & 0.61           & 1.46          & 1.67          & 94          \\ \cline{2-10} 
                         & PIP (Ours)              & \textbf{15.02} & \textbf{8.73} & \textbf{5.04} & \textbf{5.95} & \textbf{0.23}  & \textbf{0.24} & \textbf{0.12} & \textbf{16} \\ \hline
\end{tabular}%
}
\vspace{-0.3em}
\end{table*}
\begin{table*}[t]
\centering
\resizebox{\linewidth}{!}{%
\begin{tabular}{|c|c|c|c|c|c|c|c|c|c|}
\hline
\multicolumn{2}{|c|}{\multirow{2}{*}{Method}}      & \multicolumn{8}{c|}{TotalCapture}                                                                                              \\ \cline{3-10} 
\multicolumn{2}{|c|}{\multirow{2}{*}{}}            & SIP Error      & Ang Error      & Pos Error     & Mesh Error    & Rel Jitter     & Abs Jitter     & ZMP Dist      & Latency    \\ \hline
\multirow{2}{*}{Offline} & DIP~\cite{DIP}          & 18.47          & 17.54          & 9.47          & 11.19         & 2.91           & -              & -             & -          \\ \cline{2-10} 
                         & TransPose~\cite{TransPose}& 14.71          & 12.19          & 5.44          & 6.22          & 0.16           & 0.91           & 0.76          & -          \\ \hhline{|==========|}
\multirow{3}{*}{Online}  & DIP~\cite{DIP}          & 18.62          & 17.22          & 9.42          & 11.22         & 3.62           & -              & -             & 117         \\ \cline{2-10}
                         & TransPose~\cite{TransPose}& 16.58          & 12.89          & 6.55          & 7.42          & 0.95           & 1.87           & 1.40          & 94         \\ \cline{2-10}
                         & PIP (Ours)              & \textbf{12.93} & \textbf{12.04} & \textbf{5.61} & \textbf{6.51} & \textbf{0.20}  & \textbf{0.20}  & \textbf{0.23} & \textbf{16} \\ \hline
\end{tabular}%
}
\caption{
  Comparison with the state-of-the-art methods on more metrics.
  PIP outperforms previous online methods on all metrics with much less latency, while also achieves comparable capture accuracy but higher physical correctness when compared with previous offline methods.
  This demonstrates the superiority of our system which runs in real-time with extremely small latency. 
}
\label{tab:fullcmp}
\end{table*}
%
%
\section{Implementation Details}
\par
\begin{figure}
    \includegraphics[width=\linewidth]{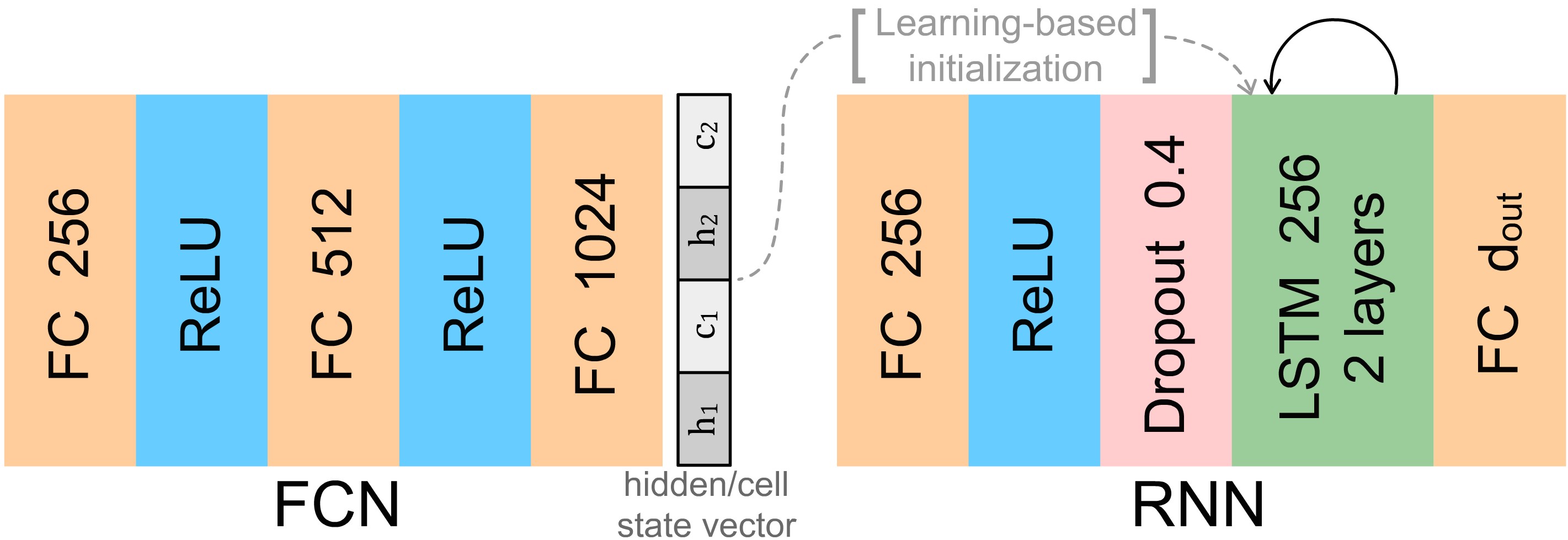}
    \caption{
        Detailed structures of the recurrent neural network (RNN) and the fully-connected network (FCN) in our kinematics module.
        "FC" represents a fully-connected layer.
        The output dimension and other hyper-parameters are marked in each block.
    }
    \label{fig:net_structure}
\end{figure}

\noindent\textbf{Network Structure.}
We schematically visualize the network structures in our kinematics module in Fig.~\ref{fig:net_structure}.
The recurrent neural network (RNN) $P_\mathrm{L}$, $P_\mathrm{A}$, $R_\mathrm{A}$, $V_\mathrm{A}$, and $C_\mathrm{F}$ share the same structure.
Each network includes a linear input layer with a ReLU activation, two Long Short-term Memory (LSTM) \cite{LSTM} layers with the width of 256, and a linear output layer.
A 40\% dropout is applied to prevent over-fitting.
The RNN $C_\mathrm{F}$ is finally activated by a Sigmoid function to obtain probability values.
The initial states of $P_\mathrm{L}$ and $V_\mathrm{A}$ are regressed from the starting leaf joint positions $\boldsymbol p_{\mathrm{leaf}}^{(0)}$ and joint velocities $\boldsymbol v^{(0)}$ using the fully-connected network (FCN) $I_\mathrm{PL}$ and $I_\mathrm{VA}$, respectively.
Each FCN consists of 3 fully-connected (FC) layers with the width of $256$, $512$, and $1024$ using the ReLU activation.
The output of the FCN is used to initialize the hidden/cell states of the two LSTM layers of the RNN at the beginning.
%
\par\noindent\textbf{Rotation Representation.} 
The inertia input vector $\boldsymbol x$ consists of accelerations and \textit{rotation matrices}, which are obtained after the calibration.
The output of $R_\mathrm{A}$ is the non-root joint rotations \textit{w.r.t} the root parameterized in the \textit{6D representation}~\cite{6D}.
Combining the estimated non-root joint rotations with the root orientation measured by the IMU placed on the pelvis, we obtain the vector $\boldsymbol \varphi$.
The character pose in the physics module is described by local joint rotations (\textit{i.e.,} each joint relative to its parent) in \textit{Euler angles}, which is denoted as $\boldsymbol \theta$.
The configuration vector $\boldsymbol q=[\boldsymbol r_\mathrm{root}\,\,\,\,\boldsymbol \theta]$ is then composed of the root translation and the pose in Euler angles.
%
%
\par
\begin{figure}
    \includegraphics[width=\linewidth]{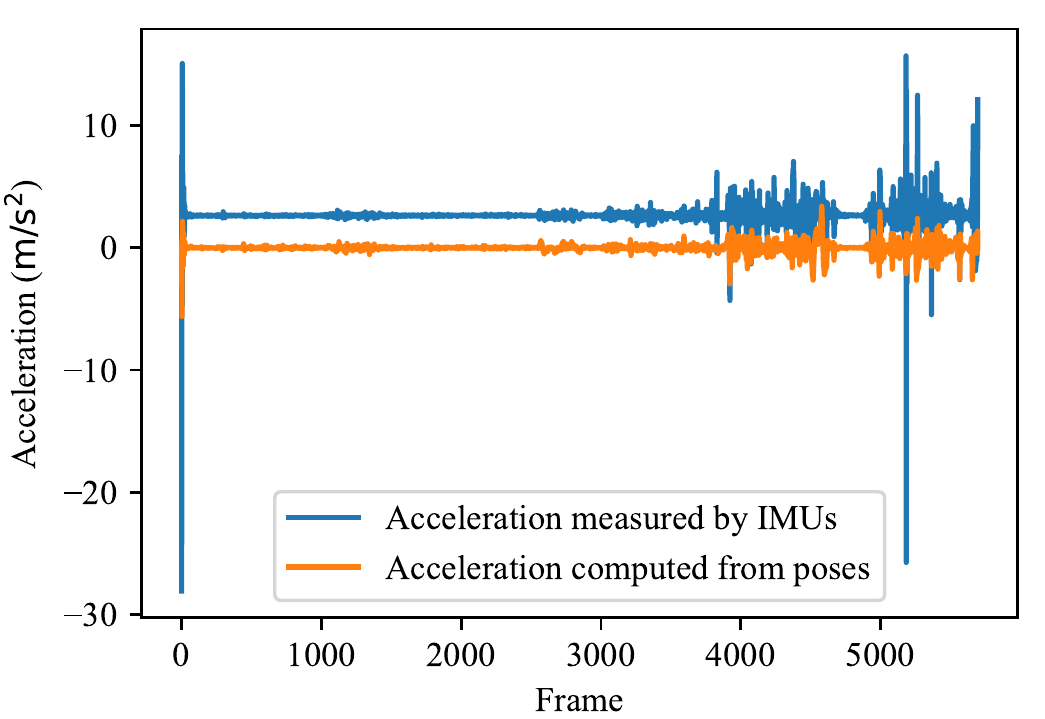}
    \caption{
        The acceleration measurements in the TotalCapture~\cite{TotalCapture} dataset is constantly biased.
        We visualize the accelerations ($x$-axis component) measured by IMUs in blue and the one computed from the subject motions based on Vicon~\cite{Vicon} by a finite-difference method in orange.
        \RED{We can see an obvious constant bias in the IMU acceleration measurements (blue) based on the fact that real accelerations should be approximately zero-centered.}
    }
    \label{fig:bias}
\end{figure}

\noindent\textbf{Datasets.}
Following~\cite{TransPose}, we use the AMASS~\cite{AMASS} dataset and the train split of the DIP-IMU~\cite{DIP} dataset for the network training, and use the TotalCapture~\cite{TotalCapture} dataset and the test split of the DIP-IMU dataset for evaluation.
For AMASS, we synthesize the IMU measurements and foot-ground contact labels as proposed by Yi et al.~\cite{TransPose}, and synthesize the ground-truth joint velocities using:
\begin{equation}
    \boldsymbol v^{\mathrm{GT}}(t) = (\boldsymbol R_\mathrm{root}^\mathrm{GT}(t))^{-1}(\boldsymbol r^\mathrm{GT}(t)-\boldsymbol r^\mathrm{GT}(t-1))/\Delta t,
\end{equation}
where $\boldsymbol R_\mathrm{root}^\mathrm{GT}(t) \in \mathbb{R}^{3\times 3}$ is the ground-truth root orientation at frame $t$; $\boldsymbol r^\mathrm{GT}\in\mathbb{R}^{3J}$ is the ground-truth joint global positions; $\Delta t$ is the frame interval.
We also re-calibrate the acceleration measurements in TotalCapture, as we find that they are constantly biased (see Fig.~\ref{fig:bias}).
%
%
Specifically, to remove the bias, we synthesize the accelerations for TotalCapture using the method of Yi et al.~\cite{TransPose} and align the mean acceleration measurement for each sequence to the mean synthetic values by adding or subtracting a constant.
%
%
%
\RED{
\par\noindent\textbf{Gain Parameters for PD Controllers.}
The gain parameters $k_{p_\theta}$, $k_{d_\theta}$, $k_{p_r}$, and $k_{d_r}$ of the dual PD controller introduced in Sec.~3.2.2 are derived as follows.
Take the joint rotation controller (controlling $\boldsymbol \theta$) as an example.
As we use first-order approximations in the dynamic states updater (Sec.~3.2.4), we apply first-order Taylor expansion on $\boldsymbol \theta$ and $\dot{\boldsymbol \theta}$, and rearrange the equation, which writes:
\begin{equation}
    \ddot{\boldsymbol \theta}(t) = \frac{1}{\Delta t^2}(\boldsymbol \theta(t+2\Delta t) - \boldsymbol \theta(t+\Delta t)) - \frac{1}{\Delta t}\dot{\boldsymbol\theta}(t),
\end{equation}
where $\Delta t=1/60$ is the time interval between frames.
By associating this equation with Eq.~3 and Eq.~5 in the main paper, the proportional gain $k_{p_\theta}$ and $k_{p_r}$ should be $3600$, and the derivative gain $k_{d_\theta}$ and $k_{d_r}$ should be $60$.
For the joint rotation controller, setting the proportional gain $k_{p_\theta}$ to a lower value gives smoother angular accelerations.
Thus, we set $k_{p_\theta}$ to $2400$ in our experiments.
}
%
%
\par\noindent\textbf{Other Details.}
We use a laptop with an Intel(R) Core(TM) i7-10750H CPU and an NVIDIA RTX2080 Super graphics card to run the experiments and the live demos.
We use PyTorch 1.8.1 with CUDA 11.1 to implement our kinematics estimator, and leverage the Rigid Body Dynamics Library~\cite{RBDL} to implement our physics-based optimizer.
The live demo is implemented using Unity3D.
We use Noitom Perception Neuron series~\cite{Noitom} IMU sensors in our demo.
Both training and evaluation assume 60 fps sensor input. 
The training data is additionally clipped into short sequences in $200$-frame lengths for more effective learning.
Specifically, we separately train each RNN in the kinematics module using the synthetic AMASS~\cite{AMASS} dataset with a batch size of 256 using the Adam~\cite{Adam} optimizer, and fine-tune $P_\mathrm{L}$ (together with $I_\mathrm{PL}$), $P_\mathrm{A}$, and $R_\mathrm{A}$ on the train split of the DIP-IMU dataset, following~\cite{TransPose}.
We do not train $V_\mathrm{A}$ and $C_\mathrm{F}$ on DIP-IMU as it does not contain global movements.
%
%
\section{Comparisons on More Metrics}
In this section, we show the comparison results with the previous state-of-the-art methods~\cite{DIP,TransPose} on more metrics.
In addition to the metrics used in the main paper, we also evaluate
\textit{1)} \textit{Angular Error}: the mean rotation error of all body joints \RED{in the global space} in degrees;
\textit{2)} \textit{Positional Error}: the mean position error of all body joints \RED{in the global space} with the root \RED{position and orientation} aligned in $\mathrm{cm}$;
\textit{3)} \textit{Relative Jitter}: the jitter calculated in the local (root-relative) frame in $\mathrm{km/s^3}$, where the root translation is not considered.
\RED{Notice that due to the length limit of the main paper, we only showed the mesh error as it incorporates both angular and positional error, and the SIP error as it is directly related to motion ambiguities in the main text.
Here, we report the results on more metrics for a fair comparison.}
We also evaluate previous offline methods for references, which need to pre-record the inertia measurements during the whole motion and estimate the motion with the help of the complete inertia sequence.
The results on TotalCapture~\cite{TotalCapture} and the test split of DIP-IMU~\cite{DIP} dataset are shown in Tab.~\ref{tab:fullcmp}.
We outperform previous online methods on all metrics with largely reduced latency, which demonstrates the accuracy and effectiveness of our approach.
Moreover, compared with the offline methods, PIP achieves comparable motion accuracy (reflected in the first 5 metrics) but higher physical plausibility (reflected in Absolute Jitter and ZMP Distance).
We attribute this to the physics-based motion optimizer proposed in the main paper.
Most importantly, our system runs \textit{in real-time}, while the offline approaches require the access to the complete inertia sequence.
Thus, our approach significantly closes the gap between online and offline methods, and enables a wide variety of real-time applications such as gaming. 

\RED{
\section{Discussions and Future Works}
\par\noindent\textbf{Quantitative Evaluations of Physics.}
A direct quantitative evaluation of physics (\textit{e.g.,} joint torques and ground reaction forces) would be advantageous. 
However, to the best of our knowledge, there is no public dataset containing both IMU measurements and ground-truth forces (either joint torques or ground reaction forces).
We believe that creating such a dataset requires research on its own, and would have great value for the community. 
For now, we can only provide qualitative visualization of torques/GRFs in our supplemental video and Fig.~5, which is intuitively plausible and in line with the references~\cite{Shahabpoor2017,Zell2020}.
Besides, as the output motion is \textit{entirely} driven by the estimated forces, the quantitative evaluation of the motion can also implicitly demonstrate the quality of our force estimation.
Furthermore, we use jitter (jerk) and ZMP distance as indirect quantitative evaluations of the physics estimation, which reflect the naturalness~\cite{Jerk} and equilibrium~\cite{ZMP} of the motion, respectively.
Since we do not adopt any explicit penalty on these two metrics, nor do we use any temporal filter or balancing technique on the motion, the better results on these two metrics actually suggest the improved physical correctness achieved by our motion optimizer.
\par
Regarding the ground contact evaluation, previous works~\cite{PhysCap,PhysAware} use mean penetration error to evaluate the non-physical foot penetration.
As we explicitly model the contacts as hard constraints, both sliding and ground penetration are \textit{strictly} avoided with any contacting part of the body.
Thus, these errors would be zero.
%
%
\begin{figure}
    \includegraphics[width=\linewidth]{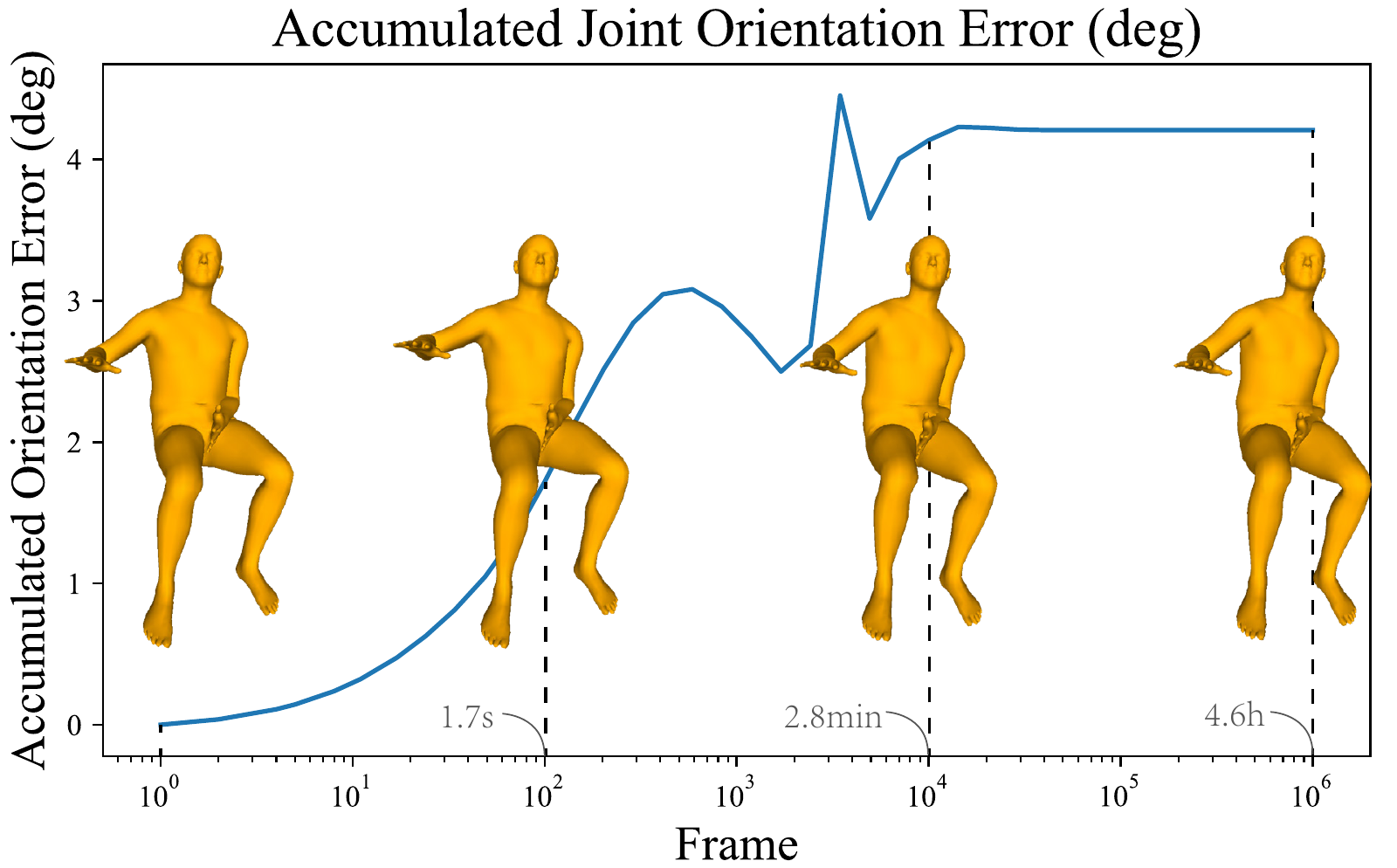}
    \caption{
        Pose drifts in a perfectly-still sitting pose.
        We evaluate PIP on 4.6-hour artificial inertia measurements with zero accelerations and fixed orientations of a sitting pose.
        We plot the accumulated orientation error of all body joints over time and pick four frames for visualization.
        Our system stably estimates a sitting pose during the entire sequence with a total drift of 4.2 degrees.
    }
    \label{fig:accumulateerror}
\end{figure}
%
%
\par\noindent\textbf{Zero Moment Point vs. Center of Pressure.}
Previous works~\cite{CoPStabilityECCV,CoPStabilityThesis} use Center of Pressure (CoP) accuracy to quantify the force estimation, which is related to our Zero Moment Point (ZMP) distance.
Here we point out the difference between these two notations and the reason why we choose to use ZMP distance.
The pressure between the human body and the ground can be represented by a force exerted at the CoP.
If such a force can balance all active forces acting on the human body during the motion, the human body is in dynamic equilibrium, and ZMP coincides with CoP (\textit{i.e.,} within the support polygon).
However, when the force acting on the CoP cannot balance other forces, the human will fall down about the foot edge, and the ZMP (more precisely, the fictitious ZMP) will be outside the support polygon, whose distance to the polygon is proportional to the intensity of the unbalanced force.
In such cases, CoP is on the border of the support polygon as the ground reaction forces cannot escape the polygon.
Thus, the reason to use ZMP distance in our physics evaluation becomes clear:
since the estimated motion cannot be perfectly physically correct and contains unbalanced movements, the ZMP distance can better reflect the disequilibrium in the captured motion.
On the other hand, evaluating CoP accuracy needs a more sophisticated modeling of human feet (rather than a simplified square facet contact) and ground-truth pressure annotations, which we leave as a future work.
For more detailed introductions of ZMP, readers are referred to~\cite{ZMP}.
%
%
\par\noindent\textbf{Drifts in Long-term Tracking.}
As a purely inertial sensor based approach, PIP inevitably suffers from drifts in long-term tracking.
As measured in Fig.~3, the translation drift of our system depends on how far the subject moves, and is about $4.6$\% in our experiments.
Regarding the subject's pose, we do not see an evident drift in our experiments.
This may be because the subject is always moving, and the orientation and acceleration measurements effectively confine the possible human pose.
Therefore, it is interesting to examine the \textit{pose drift in still poses}, especially for the ambiguous ones like sitting.
However, as the IMUs always have small noises and humans cannot keep perfectly still for a long time, it is difficult to quantify the pose drifts in real settings.   
Thus, we conduct a toy experiment where we artificially set all acceleration measurements to zero and orientations unchanged at the point after the sit-down motion in Fig.~6, \textit{i.e.,} to simulate a perfectly-still sitting pose.
As shown in Fig.~\ref{fig:accumulateerror}, our system can keep estimating sitting poses stably with a total drift of 4.2 degrees for all body joints at 1 million (4.6 hours) frames.
This demonstrates the robustness of our system in long-term tracking, which is ensured by the RNNs and the learning-based RNN initialization scheme.
We also conduct a live experiment where our method can track long-period sitting for half an hour stably and is not getting worse as time goes by. Please refer to our supplementary video for more results.
}

\end{document}